\documentclass[aps,twocolumn]{revtex4}

\usepackage{graphics}
\usepackage{epsfig}

\usepackage{amsmath}
\usepackage{amssymb}
\usepackage{stmaryrd}

\def\ov#1{\overline{#1}}

\def\vb#1{\mbox{\boldmath$#1$}}
\def\pd#1#2{\frac{\partial #1}{\partial #2}}

\def\wh#1{\widehat{#1}}
\def\bdot{\,\vb{\cdot}\,}
\def\btimes{\,\vb{\times}\,}

\def\bhat{\wh{{\sf b}}}
\def\exd{{\sf d}}

\newcommand{\bc}{\begin{center}}
\newcommand{\ec}{\end{center}}
\newcommand{\bt}{\begin{tabbing}}
\newcommand{\et}{\end{tabbing}} 
\newcommand{\be}{\begin{eqnarray*}}
\newcommand{\ee}{\end{eqnarray*}}

\begin{document}

\title{Faithful guiding-center orbits in an axisymmetric magnetic field}

\author{Alain J.~Brizard$^{1,a}$ and Brook C. Hodgeman$^{1}$}
\affiliation{$^{1}$Department of Physics, Saint Michael's College, Colchester, VT 05439, USA \\ $^{a}$Author to whom correspondence should be addressed: abrizard@smcvt.edu}

\begin{abstract}
The problem of the charged-particle motion in an axisymmetric magnetic geometry is used to assess the validity of higher-order Hamiltonian guiding-center theory, which includes higher-order corrections associated with gyrogauge invariance as well as guiding-center polarization induced by magnetic-field non-uniformity. Two axisymmetric magnetic geometries are considered: a magnetic mirror geometry and a simple tokamak geometry. When a magnetically-confined charged-particle orbit is regular (i.e., its guiding-center magnetic moment is adiabatically invariant), the guiding-center approximation, which conserves both energy and azimuthal canonical angular momentum, is shown to be faithful to the particle orbit when higher-order corrections are taken into account. 
\end{abstract}

\date{\today}

\maketitle

\section{Introduction}

The guiding-center representation \cite{Littlejohn:1983,Brizard:1989,Cary_Brizard:2009,Tronko_Brizard:2015} of charged-particle orbits is at the foundation of most particle simulations of magnetized plasmas \cite{Qin:2008,White:2014,Burby:2017,Albert:2020,Bierwage_White:2022}. The faithfulness of this representation relies on the adiabatic invariance of the magnetic moment $\mu$, which is expressed as an asymptotic expansion based on the nonuniformity of the magnetic field. In addition, when the magnetic field is axisymmetric, the connection between the exact particle canonical azimuthal angular momentum and its guiding-center representation can be used as a test for the faithfulness of the guiding-center approximation \cite{Belova:2003}.

For each guiding-center orbit parametrized by the guiding-center magnetic moment $\mu$ (with initial guiding-center position ${\bf X}_{0}$ and initial parallel guiding-center momentum $P_{\|0}$), there corresponds an infinite set of particle orbits (with local initial conditions ${\bf x}_{0}$ and $\dot{\bf x}_{0}$) that are labeled by an initial gyroangle, measured on an initial Kruskal ring \cite{Burby_Qin:2012,Burby:2020} (also parametrized by the magnetic moment $\mu$) defined on the two-dimensional plane perpendicular to the local magnetic field. (In a uniform magnetic field ${\bf B} = B\,\wh{\sf z}$, the Kruskal ring is a circle in the $(x,y)$-plane of radius $\sqrt{2\mu\,B/m\Omega^{2}}$ centered at the guiding-center position ${\bf X}_{0}$, where $\Omega = eB/mc$ denotes the gyrofrequency of a charged particle of mass $m$ and charge $e$ and $c$ denotes the speed of light.) By adopting a guiding-center formulation that is gyrogauge invariant  \cite{Littlejohn:1983,Brizard:1989} (i.e., a formulation that is not only gyroangle invariant but also independent how the gyroangle is measured), the guiding-center orbit can be compared to an arbitrary particle orbit belonging to the same initial Kruskal ring.

The purpose of our present work is to explore how higher-order Hamiltonian guiding-center theory is faithful to charged-particle dynamics in an axisymmetric magnetic field. In particular, compared to the exact particle canonical angular momentum invariant, we will investigate the faithfulness of the guiding-center canonical angular momentum derived either in the truncated (lowest-order) guiding-center model \cite{White_Chance:1984,Cary_Brizard:2009}, in the standard work of Littlejohn \cite{Littlejohn:1983,Brizard:1989}, which includes gyrogauge corrections, or the extended work of Brizard \cite{Brizard:2013} and Tronko and Brizard \cite{Tronko_Brizard:2015}, which includes effects due to guiding-center polarization \cite{Kaufman:1986}.  For this purpose, we will consider particle and guiding-center orbits in axisymmetric mirror geometry (Sec.~\ref{sec:mirror}) and in axisymmetric (simple) tokamak geometry (Sec.~\ref{sec:tokamak}).

\section{\label{sec:Lagrange}Particle and Guiding-center Orbits in General Axisymmetric Magnetic Geometry}

In the present Section, we consider the problem of charged particle motion in a generic nonuniform magnetic field. Since the primary focus of our work involves the effects of magnetic nonuniformity, we assume that the magnetic field is stationary and an electric field is absent in our formulation.

\subsection{Lagrangian particle dynamics}

The orbits of a charged particle moving in a nonuniform magnetic field ${\bf B}({\bf x})$ are represented as solutions of the Euler-Lagrange equations obtained from the particle Lagrangian
\begin{eqnarray} 
L &=& \left( \frac{e}{c}\,{\bf A}({\bf x}) \;+\; m\,{\bf v}\right)\bdot\dot{\bf x} \;-\; \frac{m}{2}\,|{\bf v}|^{2} \nonumber \\
 &=& \frac{e}{c}\,{\bf A}({\bf x})\bdot\dot{\bf x} \;+\; \frac{m}{2}\;\left|\dot{\bf x}\right|^{2},
\label{eq:Lag_particle}
\end{eqnarray}
where the particle velocity is ${\bf v} = \dot{\bf x}$ and the magnetic field ${\bf B} \equiv \nabla\btimes{\bf A} \equiv B\,\bhat$ (which is decomposed in terms of its magnitude $B = |{\bf B}|$ and its direction unit vector $\bhat = {\bf B}/|{\bf B}|$) is represented in terms of a vector potential ${\bf A}$. From this Lagrangian, we obtain the Euler-Lagrange equations
\begin{equation} 
\frac{d}{dt}\left(\pd{L}{\dot{\bf x}}\right) \;=\; \pd{L}{\bf x},
\label{eq:EL_particle}
\end{equation}
which yield the usual Lorentz force equation
\begin{equation} 
m\;\ddot{\bf x} \;=\; \frac{e}{c}\left( \nabla{\bf A}\bdot\dot{\bf x} \;-\frac{}{} 
\dot{\bf x}\bdot\nabla{\bf A} \right) \;=\; \frac{e}{c}\,\dot{\bf x}\btimes{\bf B},
\label{eq:Lorentz}
\end{equation}
which is solved subject to the initial conditions $({\bf x}_{0}, \dot{\bf x}_{0} = {\bf v}_{0})$. Since we are interested in orbital solutions of the Lorentz force equation \eqref{eq:Lorentz} over long time scales compared to the short gyration period, which is inversely proportional to the gyrofrequency $\Omega_{0} = e B_{0}/(mc)$, where $B_{0}$ denotes the characteristic strength of the magnetic field, we introduce a dimensionless time $t^{\prime} = \epsilon\,\Omega_{0}\,t$, where $\epsilon \ll 1$ denotes a small ordering parameter, so that $\dot{\bf x} = \epsilon\,\Omega_{0}\;
{\bf x}^{\prime}$ (a prime denotes a derivative with respect to $t^{\prime}$). Hence, the Lorentz force equation \eqref{eq:Lorentz} becomes
\begin{equation}
\epsilon\;\ov{\bf x}^{\prime\prime} \;=\; \ov{\bf x}^{\prime}\btimes
\ov{\bf B}(\ov{\bf x}),
\label{eq:Lorentz_epsilon}
\end{equation}
where we have introduced a characteristic length scale $R_{0}$ associated with the magnetic field, so that $\ov{\bf x} \equiv {\bf x}/R_{0}$ is dimensionless, and the magnetic field ${\bf B} = B_{0}\,\ov{\bf B}(\ov{\bf x})$ is expressed in terms of a dimensionless field $\ov{\bf B}(\ov{\bf x})$. The solution for this equation of motion, which can be carried out as an asymptotic expansion in powers of $\epsilon$ \cite{Kruskal:1958}, will be carried out numerically in this paper.

In the event the magnetic field is axisymmetric, the particle Lagrangian \eqref{eq:Lag_particle} is independent of the particle azimuthal angle $\varphi$, and the azimuthal canonical angular momentum
\begin{equation}  
P_{\varphi} \;\equiv\; \pd{L}{\dot{\varphi}} \;=\; \left( \frac{e}{c}\,{\bf A} \;+\; m\,\dot{\bf x}\right)\bdot\pd{\bf x}{\varphi} 
\label{eq:P_varphi}
\end{equation}
is a constant of the motion for particle orbits. With the normalization discussed above, we note that the azimuthal canonical angular momentum \eqref{eq:P_varphi} becomes 
\begin{equation}
\frac{P_{\varphi}}{mR_{0}^{2}\Omega_{0}} \;\equiv\; \ov{P}_{\varphi} \;=\; \left( \ov{\bf A} \;+\frac{}{} \epsilon\,\ov{\bf x}^{\prime}
\right)\bdot\pd{\ov{\bf x}}{\varphi},
\label{eq:P_varphi_epsilon}
\end{equation}
with the dimensionless magnetic vector potential $\ov{\bf A} \equiv {\bf A}/(B_{0}R_{0})$.

\subsection{Lagrangian guiding-center dynamics}

For most particle orbits that are solutions of the Lorentz force equation \eqref{eq:Lorentz}, the lowest-order magnetic moment 
\begin{equation}
\mu_{0} \equiv \frac{m}{2B}\;|\bhat\btimes\dot{\bf x}|^{2} \;=\; \frac{m}{2B}\;
|\vb{\rho}_{0}|^{2}\,\Omega^{2} 
\end{equation}
is an adiabatic invariant (where $\vb{\rho}_{0} \equiv \bhat\btimes\dot{\bf x}/\Omega$ denotes the lowest-order gyroradius), i.e., while the time derivative
\begin{equation} 
\dot{\mu}_{0} \;=\; -\,\mu_{0}\,{\bf v}\bdot\nabla\ln B - \frac{mv_{\|}}{B}\;{\bf v}
\bdot\nabla\bhat\bdot{\bf v}_{\bot} \;\neq\; 0
 \label{eq:mu0_dot}
 \end{equation}
does not vanish for a general magnetic field, its average over the fast gyromotion time scale yields
 \[ \langle\dot{\mu}_{0}\rangle = -\,\mu_{0}\,v_{\|} \left( \bhat\bdot\nabla \ln B \;+\frac{}{} \nabla\bdot\bhat\right) = -\,\frac{\mu_{0}\,v_{\|}}{B}\;
 (\nabla\bdot{\bf B}), \]
 where $v_{\|} \equiv {\bf v}\bdot\bhat$ is the local parallel velocity, $\langle{\bf v}_{\bot}\rangle = 0$ and $\langle m{\bf v}_{\bot}
 \bdot\nabla\bhat\bdot{\bf v}_{\bot}\rangle = \mu_{0} B\,(\nabla\bdot\bhat)$. Since magnetic fields are divergenceless, we immediately find that $\langle\dot{\mu}_{0}
 \rangle = 0$, i.e., $\mu_{0}$ is an invariant over time scales that are slow compared to the fast gyromotion time scale.
 
 The purpose of the guiding-center transformation is to construct an expression for the guiding-center moment 
 \begin{equation}
 \mu \;=\; \mu_{0} \;+\; \epsilon\,\mu_{1} \;+\; \cdots
 \label{eq:mu_gc}
 \end{equation}
 represented as an asymptotic series in powers of the dimensionless parameter $\epsilon$, where the first-order correction 
\begin{eqnarray}
\mu_{1} &=& \left( \mu_{0}\nabla\ln B + \frac{p_{\|}^{2}\,\vb{\kappa}}{2\,mB}\right)
\bdot\vb{\rho}_{0} - \frac{3}{2}\,\mu_{0} \left( \frac{p_{\|}\,\tau}{m\Omega}\right) \nonumber \\
 &&+\; \frac{p_{\|}}{2B}\;\frac{d\bhat}{dt}\bdot\vb{\rho}_{0}
 \label{eq:mu_1}
\end{eqnarray}
involves first-order magnetic-field nonuniformity associated with magnetic curvature $\vb{\kappa} = \bhat\bdot\nabla\bhat$ and magnetic twist $\tau = \bhat\bdot\nabla\btimes\bhat$, with $d\bhat/dt \equiv \dot{\bf x}\bdot\nabla\bhat$ in a time-independent nonuniform magnetic field. The new expression \eqref{eq:mu_1}, which is derived from the standard expression found in Refs.~\cite{Littlejohn:1983,Brizard:1989,Cary_Brizard:2009,Tronko_Brizard:2015} in App.~\ref{sec:gc}, is easily computed from the particle dynamics. Hence, from the magnetic-moment analysis of the particle orbit yields a relatively accurate value for the guiding-center magnetic moment $\mu = \mu_{0} + \epsilon\,\mu_{1}$, which can then be used as a label for the guiding-center orbit.

The reduced guiding-center representation of charged-particle dynamics in a nonuniform magnetic field \cite{Cary_Brizard:2009} is obtained by an asymptotic decoupling of the fast gyromotion from the slow magnetic-drift motion in a reduced dynamical phase space with guiding-center coordinates $Z^{\alpha} = ({\bf X},P_{\|})$, while the fast gyromotion is represented by the canonically-conjugate guiding-center action-angle coordinates $(J,\zeta)$, where the gyroaction $J \equiv \mu B_{0}/\Omega_{0}$ (defined in terms of the magnetic moment $\mu$) is canonically conjugate to the guiding-center gyroangle $\zeta$.

The guiding-center Lagrangian is expressed up to first order in magnetic-field nonuniformity as
\begin{eqnarray} 
L_{\rm gc} &=& \left( \frac{e}{\epsilon\,c}\,{\bf A}({\bf X}) \;+\; P_{\|}\,\bhat({\bf X}) \;-\; \epsilon\,J\;
\vb{\cal R}^{*}({\bf X})\right)\bdot\dot{\bf X} \nonumber \\
 &&+\; \epsilon\,J\;\dot{\zeta} \;-\; \left( \frac{P_{\|}^{2}}{2m} \;+\; \mu B({\bf X})\right),
\label{eq:Lag_gc}
\end{eqnarray}
where the $\epsilon$-ordering introduced in Eq.~\eqref{eq:Lag_gc} is based on the standard {\it macroscopic} ordering associated with the renormalization of the electric charge $e \rightarrow e/\epsilon$. The selection of the vector field
\begin{equation}
 \vb{\cal R}^{*}({\bf X}) \;=\; \left\{ \begin{array}{lr}
 0 & (\mbox{A}) \\
  \vb{\cal R} \;+\; \frac{1}{2}\,\tau\,\bhat & (\mbox{B})  \\
 \vb{\cal R} \;+\; \frac{1}{2}\,\nabla\btimes\bhat & (\mbox{C}) 
 \end{array} \right.
 \label{eq:R_star_def}
\end{equation} 
is based on whether the gyrogauge vector field $\vb{\cal R} \equiv \nabla\wh{\sf e}_{1}\bdot\wh{\sf e}_{2}$ (defined \cite{Littlejohn:1983} in terms of the local orthogonal unit vectors $\wh{\sf e}_{1}$ and $\wh{\sf e}_{2} \equiv \bhat\btimes\wh{\sf e}_{1}$) is kept (B \& C) or not (A), and whether the guiding-center polarization correction  \cite{Tronko_Brizard:2015} is kept (C) or not (B). The {\it truncated} guiding-center model (A) is the simplest guiding-center model that is used in several guiding-center orbit codes (e.g., \cite{White_Chance:1984,White:2014})  and is reviewed in Ref.~\cite{Cary_Brizard:2009}. The {\it standard} guiding-center model (B) was derived \cite{Littlejohn:1983,Brizard:1989} to ensure that the guiding-center equations of motion are not only independent of the gyroangle but also independent of how this gyroangle is locally measured in the perpendicular plane spanned by the unit vectors $(\wh{\sf e}_{1}, \wh{\sf e}_{2})$.  The {\it extended} guiding-center model (C) was derived \cite{Tronko_Brizard:2015} to ensure that the guiding-center transformation accurately represents the guiding-center polarization \cite{Kaufman:1986,Brizard:2013}.

The guiding-center Euler-Lagrange equations 
\[ d(\partial L_{\rm gc}/\partial\dot{Z}^{\alpha})/dt \;=\; \partial L_{\rm gc}/\partial Z^{\alpha} \]
lead to the reduced guiding-center equations of motion
\begin{eqnarray}
\dot{\bf X} &=& \frac{P_{\|}}{m}\;\frac{{\bf B}^{*}}{B_{\|}^{*}} \;+\; \frac{\epsilon\,c\bhat}{eB_{\|}^{*}}\btimes \mu\,\nabla B, \label{eq:Xgc_dot} \\
\dot{P}_{\|} &=& -\;\frac{{\bf B}^{*}}{B_{\|}^{*}}\bdot \mu\,\nabla B, \label{eq:Pgc_dot}
\end{eqnarray}
where
\begin{equation}
{\bf B}^{*} \;=\; {\bf B} \;+\; \frac{\epsilon\,c}{e}\left(P_{\|}\;\nabla\btimes\bhat \;-\frac{}{}
\epsilon\,J\;\nabla\btimes\vb{\cal R}^{*}\right), \label{eq:B_star}
\end{equation}
and $B_{\|}^{*} \equiv \bhat\bdot{\bf B}^{*}$ can be used as the guiding-center Jacobian. These guiding-center equations are solved subject to the initial conditions $({\bf X}_{0},P_{\|0})$, once again labeled by the guiding-center magnetic moment $\mu$ obtained from the particle orbit. Because the particle and guiding-center orbits share the same values of energy $E$ and magnetic moment $\mu$, the initial guiding-center parallel momentum can be chosen from the initial condition $P^{2}_{\|0}/2m = E - \mu\,
B({\bf X}_{0})$, where the initial guiding-center position ${\bf X}_{0}$ is calculated from the initial particle position ${\bf x}_{0}$ according to the guiding-center transformation \cite{Littlejohn:1983,Brizard:1989}
\begin{equation}
{\bf X} \;=\; {\bf x} + \epsilon\,G_{1}^{{\bf x}} + \epsilon^{2}\,G_{2}^{{\bf x}} + \frac{1}{2}\,\epsilon^{2}\,{\sf G}_{1}\cdot\exd G_{1}^{{\bf x}} + \cdots,
\label{eq:initial_X}
\end{equation}
which implies that the initial guiding-center position ${\bf X}_{0}$ is shifted from the initial particle position ${\bf x}_{0}$. Hence, the initial guiding-center position ${\bf X}_{0}$ will depend on the guiding-center model used in Eq.~\eqref{eq:R_star_def}, which differs at second order through $G_{2}^{{\bf x}}$ \cite{Tronko_Brizard:2015}.

Lastly, when the magnetic field is axisymmetric, the guiding-center Lagrangian \eqref{eq:Lag_gc}  is independent of the guiding-center azimuthal angle $\Phi$, and the guiding-center azimuthal canonical angular momentum
\begin{equation}
P_{{\rm gc}\Phi} = \left( \frac{e{\bf A}}{\epsilon\,c} + P_{\|}\,\bhat - \epsilon\,J\;\vb{\cal R}^{*}\right)\bdot\pd{\bf X}{\Phi} \equiv \frac{e{\bf A}^{*}}{\epsilon c}\bdot\pd{\bf X}{\Phi}
\label{eq:Pgc_varphi}
\end{equation}
is an exact guiding-center invariant. We note that the terms of third order in $\epsilon^{3}$ in ${\bf A}^{*}$ (i.e., second order in magnetic-field nonuniformity) and higher are ignored.

\subsection{Validity of the guiding-center representation in general axisymmetric magnetic geometry}

While the azimuthal canonical angular momenta \eqref{eq:P_varphi} and \eqref{eq:Pgc_varphi} are constants of motion of their respective equations of motion, 
they can only be compared when the guiding-center azimuthal canonical angular momentum \eqref{eq:Pgc_varphi} is pulled back into particle phase space:
\begin{eqnarray} 
{\sf T}_{\rm gc}P_{{\rm gc}\Phi} &=& P_{{\rm gc}\Phi} + \epsilon\left(G_{1}^{\mu}\pd{}{\mu} + G_{1}^{p_{\|}}\pd{}{p_{\|}} - \vb{\rho}_{0}\bdot\nabla\right)P_{{\rm gc}\Phi} \nonumber \\ 
 &&+\; \epsilon^{2} \left(\frac{1}{2}\,\vb{\rho}_{0}\vb{\rho}_{0}:\nabla\nabla - \vb{\rho}_{1}\vb{\cdot}\nabla\right)P_{{\rm gc}\Phi} + \cdots \nonumber \\
  &\equiv& P_{\varphi},
\label{eq:Tgc_Pgc}
\end{eqnarray} 
where $(G_{1}^{\mu}, G_{1}^{p_{\|}})$ denote the first-order corrections to the guiding-center magnetic moment and guiding-center parallel momentum, respectively, and the first-order gyroradius is defined in particle phase space as
\begin{equation}
\vb{\rho}_{1} \;=\; -\;\frac{\mu_{0}B\,\vb{\kappa}}{2m\Omega^{2}} + \left(\frac{\vb{\rho}_{0}}{2}\bdot\nabla\ln B - \frac{p_{\|}\tau}{m\Omega}\right)\vb{\rho}_{0} + \rho_{1\|}\,\bhat,
\label{eq:rho_1}
\end{equation}
where, while an explicit expression for the parallel component $\rho_{1\|} \equiv \vb{\rho}_{1}\bdot\bhat$ will not be needed in the present work, we note that its gyroangle average is $\langle\rho_{1\|}\rangle = (\mu B/2m\Omega^{2})\nabla\bdot\bhat$. We note that the first term in Eq.~\eqref{eq:rho_1} appears as a result of the extension \cite{Tronko_Brizard:2015} of the standard guiding-center transformation \cite{Littlejohn:1983} that correctly calculates the guiding-center polarization (see App.~\ref{sec:gc_pol} for details).

The guiding-center representation is faithful to particle dynamics if the identity \eqref{eq:Tgc_Pgc} is satisfied up to an arbitrary order in $\epsilon$. This identity guarantees that the guiding-center push-forward of the particle conservation law $dP_{\varphi}/dt = 0$ yields the guiding-center conservation law
\[
0 = {\sf T}_{\rm gc}^{-1}\left(\frac{dP_{\varphi}}{dt}\right) =  \left[{\sf T}_{\rm gc}^{-1}\left(\frac{d}{dt}{\sf T}_{\rm gc}\right)\right]{\sf T}_{\rm gc}^{-1}P_{\varphi} \equiv \frac{d_{\rm gc}P_{{\rm gc}\Phi}}{dt},
\]
where $d_{\rm gc}/dt$ is the time derivative generated by the guiding-center Lagrangian dynamics and the guiding-center invariant $P_{{\rm gc}\Phi} \equiv {\sf T}_{\rm gc}^{-1}P_{\varphi}$ may be truncated at an arbitrary order in $\epsilon$.
 
In previous work, Belova {\it et al.} \cite{Belova:2003} considered energetic-particle orbits in the National Spherical Torus Experiment (NSTX) that satisfied the adiabatic invariance of the higher-order guiding-center magnetic moment \eqref{eq:mu_gc}, and numerically investigated the validity of the guiding-center representation by verifying that the explicit expression for the guiding-center pull-back of the guiding-center toroidal canonical angular momentum
\begin{equation} 
{\sf T}_{\rm gc}P_{{\rm gc}\Phi} = P_{{\rm gc}\Phi}\left({\bf x} - \epsilon\vb{\rho}, p_{0\|} + \epsilon\,G_{1}^{p_{\|}}, \mu_{0} + \epsilon G_{1}^{\mu}\right)
\label{eq:Pphi_Belova}
\end{equation}
is nearly invariant, where $\vb{\rho} = {\bf x} - {\sf T}_{\rm gc}{\bf X}$ includes the first-order corrections \eqref{eq:rho_1} due to magnetic nonuniformity. In Fig.~3 of Ref.~\cite{Belova:2003}, Belova {\it et al.} \cite{Belova:2003} show that the standard guiding-center expression $\psi^{*} = \psi - \epsilon\,(cP_{\|}/e)b_{\Phi} + \epsilon^{2} (cJ/e){\cal R}_{\Phi}^{*}$, where ${\cal R}_{\Phi}^{*} = b_{z} + \frac{1}{2}\tau b_{\Phi}$, yields an improved particle canonical angular momentum invariant \eqref{eq:Pphi_Belova} compared with the truncated guiding-center expression $\psi^{*} = \psi - \epsilon\,(cP_{\|}/e)b_{\Phi}$.

In the present paper, the validity of the guiding-center representation is assessed on the basis of verifying that the expansion \eqref{eq:Tgc_Pgc} is exactly valid at each order in $\epsilon$ for two axisymmetric magnetic geometries: mirror geometry 
(Sec.~\ref{sec:mirror}) and simple tokamak geometry (Sec.~\ref{sec:tokamak}). Here, the guiding-center representation is shown to be faithful up to (and including) first order in magnetic-field nonuniformity, which requires retaining all terms appearing in Eq.~\eqref{eq:Pgc_varphi}. 

\section{\label{sec:mirror}Magnetic Mirror Geometry}

\begin{figure}
\epsfysize=2.5in
\epsfbox{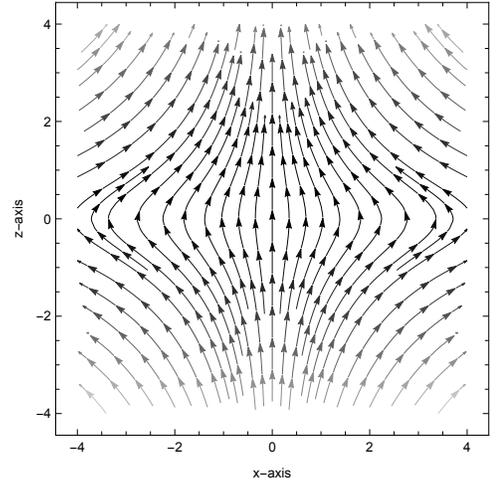}
\caption{Plot of the magnetic-mirror vector field \eqref{eq:B_mirror} in the $(x,z)$-plane.}
\label{fig:B_mirror}
\end{figure}

We begin by considering the problem of charged-particle motion in axisymmetric magnetic mirror geometry, where the magnetic field is represented, using the cylindrical coordinates $(r,\varphi,z)$, by the dimensionless expression
\begin{eqnarray}
{\bf B} &=& \nabla\psi(r,z)\btimes\nabla\varphi \;=\; \frac{1}{r}\left( \pd{\psi}{r}\;
\wh{\sf z} \;-\; \pd{\psi}{z}\;\wh{\sf r}\right) \nonumber \\
 &=& B_{r}(r,z)\;\wh{\sf r} \;+\; B_{z}(z)\;\wh{\sf z},
\label{eq:B_mirror}
\end{eqnarray}
where the magnetic flux $\psi(r,z)$ is defined as
\begin{equation}
\psi(r,z) \;=\; \frac{1}{2}\;r^{2}\,(1 + z^{2}).
\label{eq:psi_def}
\end{equation}
Figure \ref{fig:B_mirror} shows the magnetic-mirror vector field \eqref{eq:B_mirror} in the $(x,z)$-plane. With a field line passing through the point $r = r_{0}$ on the equatorial plane $z = 0$, which is labeled by the magnetic flux $\psi(r_{0},0) = r_{0}^{2}/2$, the radial coordinate $r(z) = r_{0}/\sqrt{1 + z^{2}}$ of the field line can be expressed as a function of $z$.

The magnitude of the magnetic field \eqref{eq:B_mirror} is 
\begin{equation}
B(r,z) \;=\; \sqrt{r^{2}\,z^{2} \;+\; (1 + z^{2})^{2}},
\label{eq:mag_mirror}
\end{equation}
while the unit vector along the magnetic field is
\begin{equation}
\bhat(r,\varphi,z) \;=\;  b_{r}(r,z)\;\wh{\sf r}(\varphi) \;+\; b_{z}(r,z)\;\wh{\sf z}
\label{eq:unit_mirror}
\end{equation}
where $b_{r}(r,z) = -\,r\,z/B(r,z)$, $b_{z}(r,z) = (1 + z^{2})/B(r,z)$, and $\wh{\sf r}(\varphi) = \cos\varphi\;\wh{\sf x} + \sin\varphi\;\wh{\sf y}$. Next, we can calculate
\begin{eqnarray}
\bhat\btimes\nabla\ln B &=& -\frac{r\,b_{r}}{B} \left(3\,z b_{z} - r b_{r}\right)
\nabla\varphi \equiv -\,K_{1}\nabla\varphi, \label{eq:G_mirror} \\
\nabla\btimes\bhat &=& -\,\left( r^{2}/B + K_{1}\right)\nabla\varphi \equiv -\,K_{2}\nabla\varphi, \label{eq:K_mirror}
\end{eqnarray}
with $\nabla\btimes{\bf B} = -r^{2}\,\nabla\varphi$.

\subsection{Particle dynamics}

The normalized equations of motion \eqref{eq:Lorentz_epsilon} are expressed in cylindrical coordinates as
\begin{equation} 
\epsilon\,{\bf x}^{\prime\prime} = r\,\varphi^{\prime} \left( B_{z}\,\wh{r} - B_{r}\,\wh{\sf z}\right) - \left( B_{z}\,r^{\prime} - B_{r}\,z^{\prime}
\right)\wh{\varphi},
\label{eq:motion_mirror}
\end{equation}
where ${\bf x}^{\prime} = r^{\prime}\,\wh{r} + r\,\varphi^{\prime}\,\wh{\varphi} + z^{\prime}\,\wh{\sf z}$, with $\wh{\varphi} = \partial\wh{r}/\partial\varphi$. These dimensionless equations are numerically solved for $\epsilon = 1/20$, with the initial conditions $(r_{0},\varphi_{0},z_{0}) = (1,0,0)$ and $(r_{0}^{\prime},\varphi_{0}^{\prime},z_{0}^{\prime}) = (0, 1/10,\sqrt{24}/10)$ associated with the dimensionless energy $E = 1/4$. 

We note that, because of the azimuthal symmetry of the magnetic field \eqref{eq:B_mirror}, i.e., the components $(B_{r},B_{z})$ are independent of the azimuthal angle $\varphi$, the azimuthal canonical angular momentum
\begin{equation} 
P_{\varphi} \;=\; \frac{1}{\epsilon}\,\psi(r,z) \;+\; r^{2}\varphi^{\prime}
\label{eq:P_theta}
\end{equation}
is a constant of the motion.

\begin{figure}
\epsfysize=2in
\epsfbox{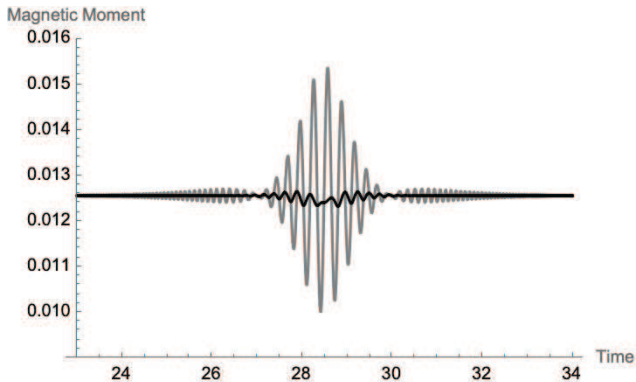}
\caption{Plots of the normalized lowest-order magnetic moment $\mu_{0}$ (gray) and the normalized magnetic moment $\mu = \mu_{0} + \epsilon\,\mu_{1} \simeq 0.01257$ (black), with first-order correction $\mu_{1}$ given by Eq.~\eqref{eq:mu1_mirror}. Here, 
$\epsilon = 1/20$ and the normalized kinetic energy is $E_{0} = 1/4$, with the initial conditions $(r_{0},\varphi_{0},z_{0}) = (1,0,0)$ and $(r_{0}^{\prime},\varphi_{0}^{\prime},z_{0}^{\prime}) = (0,1/10,\sqrt{24}/10)$.}
\label{fig:mu_mirror}
\end{figure}

Figure \ref{fig:mu_mirror} shows that, while the lowest-order normalized magnetic moment (normalized to $\epsilon^{2}m\Omega_{0}^{2}/2B_{0}$) 
\begin{equation} 
\mu_{0} \;=\; \left[ \left(r\,\varphi^{\prime}\right)^{2} \;+\;
\left( b_{z}\,r^{\prime} \;-\; b_{r}\,z^{\prime}\right)^{2}\right]/B
\label{eq:mu0_mirror}
\end{equation}
is relatively well conserved when the particle is near the orbital bounce points, its adiabatic invariance is compromised as the particle crosses the equatorial plane $(z = 0)$ between $t' = 28$ and 29. The addition of the first-order correction 
\begin{equation}
\mu_{1} \;=\; \left( \mu_{0}\;K_{1} \;+\; \ov{p}_{\|}^{2}\;K_{2}/B \;+\;
\ov{p}_{\|}z^{\prime}\;K_{3} \right)\varphi^{\prime}/B
\label{eq:mu1_mirror}
\end{equation} 
computed from Eq.~\eqref{eq:mu_1} (with $\tau = \bhat\bdot\nabla\btimes\bhat = 0$), where $\ov{p}_{\|} = b_{r}\,r^{\prime} + b_{z}\,z^{\prime}$ and
\begin{equation}
\begin{array}{rcl}
\vb{\rho}_{0}\bdot\nabla\ln B &=& \epsilon\,\varphi^{\prime}K_{1}/B, \\
\vb{\rho}_{0}\bdot(\bhat\bdot\nabla\bhat) &=& \epsilon\,\varphi^{\prime}K_{2}/B, \\
\vb{\rho}_{0}\bdot \bhat^{\prime} &=& \epsilon\,z^{\prime}\varphi^{\prime}K_{3},
\end{array}
\end{equation}
greatly improves the adiabatic invariance of the magnetic moment and validates the guiding-center representation for the particle orbits in the magnetic-mirror vector field \eqref{eq:B_mirror}. Here, $K_{1}$ and $K_{2}$ are defined in Eqs.~\eqref{eq:G_mirror}-\eqref{eq:K_mirror} and $K_{3} = r (3z b_{r} + r b_{z})/B^{2}$. Hence, the adiabatic invariance of the magnetic moment justifies our use of the guiding-center approximation in describing particle orbits in magnetic mirror geometry. The numerical value $\mu \simeq 0.01257$ will be used as the normalized guiding-center magnetic moment in the guiding-center equations of motion in magnetic mirror geometry.

\subsection{Guiding-center dynamics}

The guiding-center Lagrangian in magnetic mirror geometry is
\begin{equation}
L_{\rm gc} \;=\; \frac{1}{\epsilon}\psi^{*}\,\Phi^{\prime} + P_{\|}\bhat\bdot{\bf X}^{\prime} \;-\;
\left( \frac{1}{2}\,P_{\|}^{2} + J\,B\right),
\label{eq:Lgc_mirror}
\end{equation}
where we use the extended guiding-center model (C) in Eq.~\eqref{eq:R_star_def}, ${\bf X} = (R,\Phi,Z)$ denote the guiding-center position in cylindrical geometry, $P_{\|}$ denotes the normalized guiding-center parallel momentum, and $J \equiv \mu/2$ denotes the normalized guiding-center gyroaction. In addition, in magnetic mirror geometry, we may choose the perpendicular unit vectors $\wh{\sf e}_{1} = \wh{\Phi}$ and $\wh{\sf e}_{2} = \bhat\btimes\wh{\Phi}$, so that the gyrogauge vector $\vb{\cal R} = \nabla\wh{\sf e}_{1}\bdot\wh{\sf e}_{2} = b_{z}\,\nabla\Phi$ and the effective magnetic flux
\begin{equation}
    \psi^{*} \;\equiv{}\; \psi(R,Z) \;-\; \epsilon^{2}J\,\left(b_{z} - \frac{1}{2}\,K_{2}\right)
\end{equation}
is expressed in terms of the extended guiding-center model (C) in Eq.~\eqref{eq:R_star_def}: 
\begin{equation}
\vb{\cal R}^{*} \;=\; \left(b_{z} - \frac{1}{2}K_{2}\right)\nabla\Phi. 
\label{eq:Rstar_mirror}
\end{equation}
From this Lagrangian, we obtain Euler-Lagrange equations that can be expressed as Eqs.~\eqref{eq:Xgc_dot}-\eqref{eq:Pgc_dot}, where
\begin{eqnarray}
    {\bf B}^{*} &=& \nabla\psi^{*}\btimes\nabla\Phi \;-\; \epsilon\,P_{\|}\;K_{2}\,\nabla\Phi, \nonumber \\
     && \\
    B_{\|}^{*} &=& \bhat\bdot\nabla\psi^{*}\btimes\nabla\Phi, \nonumber
\end{eqnarray}
The guiding-center equations are thus expressed as
\begin{eqnarray}
(\dot{R},\;\dot{Z}) &=& \frac{P_{\|}}{RB_{\|}^{*}}\left(-\; \pd{\psi^{*}}{Z},\pd{\psi^{*}}{R}\right), \label{eq:RZ_gc} \\
\dot{\Phi} &=& -\;\frac{\epsilon}{R^{2}B_{\|}^{*}} \left( P_{\|}^{2}\;K_{2} \;+\; J\,B\;K_{1}\right), \label{eq:Phi_gc} \\
\dot{P}_{\|} &=& -\;\frac{J}{RB_{\|}^{*}}\left(\pd{B}{Z}\,\pd{\psi^{*}}{R} - \pd{B}{R}\,\pd{\psi^{*}}{Z}\right), \label{eq:P_gc}
\end{eqnarray}
which exactly conserve the (dimensionless) guiding-center energy ${\cal E} = P_{\|}^{2}/2 + J B$ (where $J = \mu/2$ is obtained from Fig.~\ref{fig:mu_mirror}) and the guiding-center canonical azimuthal angular momentum $P_{{\rm gc}\Phi} = \epsilon^{-1}\psi^{*}$. These dimensionless guiding-center equations are solved for $\epsilon = 1/20$ with the initial conditions $(R_{0},\Phi_{0},Z_{0}) = (201/200, 0, 0)$, which takes into account the radial shift \eqref{eq:initial_X} from the initial particle position $(r_{0},\varphi_{0},z_{0}) = (1,0,0)$, and $P_{\|0} = \sqrt{E - J\,B(R_{0},Z_{0})}$.

\begin{figure}
\epsfysize=1.8in
\epsfbox{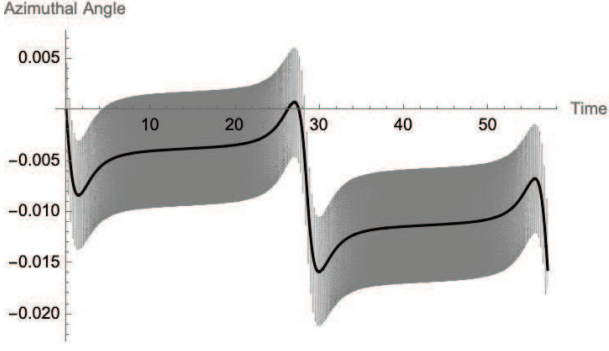}
\caption{Plots of the particle azimuthal angle $\varphi$ (gray) and the guiding-center azimuthal angle $\Phi$ (black) during a bounce period.}
\label{fig:theta_gc_mirror}
\end{figure}

Figure \ref{fig:theta_gc_mirror} shows shows plots of the particle azimuthal angle $\varphi$ (gray) and the guiding-center azimuthal angle $\Phi$ (black) for a particle orbit in the magnetic-mirror vector field \eqref{eq:B_mirror} during a full bounce period. These orbital solutions are obtained by numerical integration of the normalized equations of motion \eqref{eq:motion_mirror} and \eqref{eq:RZ_gc}-\eqref{eq:P_gc}, while conserving energy and azimuthal canonical angular momentum within machine precision. We note that the guiding-center azimuthal angle changes very slowly when the particle is near a turning point, while it changes rapidly as the particle crosses the equatorial plane ($z = 0$).

\begin{figure}
\epsfysize=3.5in
\epsfbox{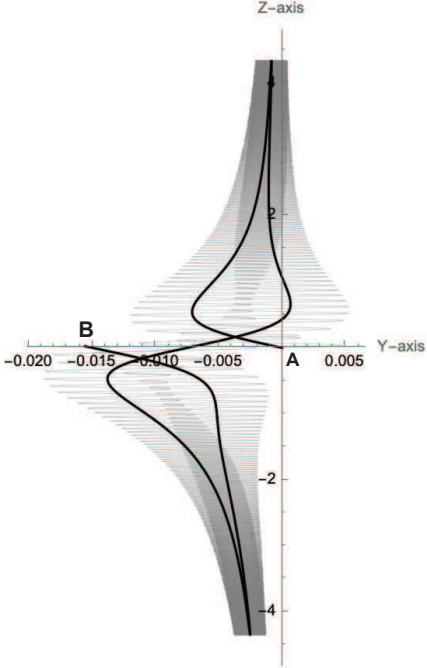}
\caption{Plots of particle orbit (gray) and guiding-center orbit (black) in the $(y,z)$ plane during a bounce period. Here, the particle and guiding-center orbits begin at point A and, after one bounce period, the guiding-center orbit has drifted (in the negative-y direction) to point B.}
\label{fig:yz_gc_mirror}
\end{figure}

Figure \ref{fig:yz_gc_mirror}, on the other hand, shows plots of the particle orbit (gray) and the guiding-center orbit (black) during a bounce period in the $(y,z)$-plane. While the motion is periodic in $(r,z)$, there is a slow drift motion in the azimuthal direction, which can be seen in Figs.~\ref{fig:theta_gc_mirror} and \ref{fig:yz_gc_mirror} (the slow drift motion is proceeds to the left on the $y$-axis). The rapid, small-amplitude oscillations that are noticeable in Figs.~\ref{fig:theta_gc_mirror} and \ref{fig:yz_gc_mirror} are due to the fast gyromotion of a charged particle about a magnetic-field line. 

\subsection{Validity of the guiding-center representation in mirror geometry}

We now show that the guiding-center representation of particle motion in magnetic mirror geometry is faithful to the exact particle motion, by showing that the guiding-center pull-back ${\sf T}_{\rm gc}P_{{\rm gc}\Phi} = P_{\varphi}$ of the guiding-center canonical azimuthal angular momentum is equal to the particle canonical azimuthal angular momentum. 

Up to second order in $\epsilon$, the guiding-center pull-back ${\sf T}_{\rm gc}P_{{\rm gc}\Phi}$ is expressed as
\begin{eqnarray}
{\sf T}_{\rm gc}P_{{\rm gc}\Phi} &=& \frac{1}{\epsilon} \left( \psi - \epsilon\vb{\rho}_{0}\bdot
\nabla\psi - \epsilon^{2}\vb{\rho}_{1}\bdot\nabla\psi \right) \nonumber \\
 &&+\; \frac{\epsilon}{2}\;\vb{\rho}_{0}\vb{\rho}_{0}:\nabla\nabla\psi  - \epsilon\,J \left( b_{z} - \frac{1}{2}\,K_{2}\right),
 \label{eq:Tgc_Pphi_mirror}
\end{eqnarray}
where the first-order gyroradius correction is given by Eq.~\eqref{eq:rho_1} and the contribution associated with the magnetic twist $\bhat\bdot\nabla\btimes\bhat = 0$ vanishes for magnetic mirror geometry. Here, we find $\vb{\rho}_{0}\bdot\nabla\psi = -\,r^{2}\varphi^{\prime}$, $\vb{\kappa}\bdot\nabla\psi = 
-\,K_{2}B$, and $\vb{\rho}_{0}\bdot\nabla\ln B = K_{1}\,\varphi^{\prime}/B$, so that
\[ -\,\vb{\rho}_{1}\bdot\nabla\psi \;=\; -\,\frac{1}{2}\,J\,K_{2} \;+\; \frac{r^{2}\varphi^{\prime 2}}{2\;B}\,K_{1}, \]
while
\[ \frac{1}{2}\,\vb{\rho}_{0}\vb{\rho}_{0}:\nabla\nabla\psi \;=\; J\,b_{z} \;-\; \frac{r^{2}\varphi^{\prime 2}}{2\;B}\,K_{1}. \]
Hence, up to second order in $\epsilon$, the guiding-center pull-back \eqref{eq:Tgc_Pphi_mirror} yields
\begin{equation}
{\sf T}_{\rm gc}P_{{\rm gc}\Phi} \;=\; \frac{1}{\epsilon}\,\psi \;+\; r^{2}\varphi^{\prime} \;\equiv\; P_{\varphi},
\end{equation}
which confirms the validity of the guiding-center representation in magnetic mirror geometry.

\section{\label{sec:tokamak}Simple Tokamak Magnetic Geometry}

We now turn our attention to the problem of charged-particle motion in a simple magnetic tokamak geometry, with circular concentric magnetic surfaces without Shafranov shift. The magnetic field is represented, using the quasi-cylindrical coordinates $(r,\vartheta,\varphi)$, by the dimensionless expression
\begin{eqnarray}
{\bf B} \;\equiv\; \frac{r^2}{qh} \,\nabla \vartheta \;+\; \nabla \varphi,
\label{eq:B_tokamak}
\end{eqnarray}
where $q(r) = q_0 + \sigma r^2/2$ is the safety factor (we will use $q_{0} = 1$ and $\sigma = 2$ in our numerical calculations) and $h = 1 + r \cos{\vartheta}$ is the normalized distance from the magnetic axis $(r = 0)$ to the particle position in the poloidal plane (which  is normalized by the major radius $R_{0}$ of the magnetic axis). Since the magnetic field \eqref{eq:B_tokamak} is divergenceless 
\begin{equation}
{\bf B} \equiv \nabla\btimes{\bf A} \;=\; \nabla\Psi\btimes\nabla\vartheta + \nabla\varphi\btimes\nabla\psi,
\end{equation}
it can be written in terms of the vector potential
\begin{equation}
{\bf A} \;=\;  \Psi \, \nabla \vartheta \;-\; \psi \,\nabla \varphi.
\label{eq:A}
\end{equation}
By comparing with Eq.~\eqref{eq:B_tokamak}, we find $\partial\Psi(r,\vartheta)/\partial r = r/h(r,\vartheta)$ and $d\psi/dr = r/q(r)$, so that the toroidal and poloidal magnetic fluxes $\Psi$ and $\psi$ are
\begin{eqnarray}
\Psi(r,\vartheta) &=& \int_{0}^{r}\frac{u \, du}{h(u , \vartheta)} = \frac{1}{\cos{\vartheta}} \left( r - \frac{\ln h(r,\vartheta)}{\cos\vartheta} \right), \label{eq:Psi_def} \\
\psi(r) &=& \int_{0}^{r}\frac{u \, du}{q(u)} = \frac{1}{\sigma} \, \ln\left(\frac{q(r)}{q_0}\right), \label{eq:psi_def}
\end{eqnarray}
where we chose $\Psi(0,\vartheta) = 0 = \psi(0)$. In what follows, we use the quasi-cylindrical unit vectors $\wh{\sf r} = \cos\vartheta\,\wh{\rho} + \sin\vartheta\,\wh{\sf z}$, $\wh{\vartheta} = \partial\wh{\sf r}/\partial\vartheta =  -\,\sin\vartheta\,\wh{\rho} + \cos\vartheta\,\wh{\sf z}$, and $\wh{\varphi} \equiv \wh{\sf r}\btimes\wh{\vartheta} = \wh{\rho}\btimes\wh{\sf z} = \partial\wh{\rho}/\partial\varphi$, with the quasi-cylindrical Jacobian ${\cal J} = (\nabla r\btimes\nabla\vartheta\bdot\nabla\varphi)^{-1} = r\,h(r,\vartheta)$.

The magnitude of the magnetic field \eqref{eq:B_tokamak} is 
\begin{equation}
B(r,\vartheta) \;=\; \frac{1}{h(r,\vartheta)}\sqrt{1 + r^2 /q(r)^2} \;\equiv\; \frac{\beta(r)}{h(r,\vartheta)}, 
\label{eq:mag_tokamak}
\end{equation}
and the unit vector along the magnetic field is
\begin{equation}
\bhat(r,\vartheta) = \frac{r}{q\beta}\;\wh{\vartheta} \;+\; \frac{1}{\beta}\;\wh{\varphi} \;\equiv\; b_{\vartheta}(r) \, \nabla \vartheta \;+\; b_{\varphi}(r,\vartheta) \, \nabla \varphi,
\label{eq:unit_tokamak}
\end{equation}
with $\nabla\btimes\bhat = \nabla b_{\vartheta}\btimes\nabla\vartheta + \nabla b_{\varphi}\btimes\nabla\varphi$. Next, we can calculate
\begin{eqnarray}
\nabla\ln B &=& \left(\frac{r\,g}{q\beta^{2}} - \frac{\cos\vartheta}{h}\right)\;\wh{\sf r} + \frac{\sin\vartheta}{h}\;\wh{\vartheta}, \\
\vb{\kappa} &=& \frac{\sin\vartheta}{h\beta^{2}}\left(\wh{\vartheta} - \frac{r}{q}\wh{\varphi}\right) - \frac{\wh{\sf r}}{\beta^{2}}\left(\frac{\cos\vartheta}{h} + \frac{r}{q^{2}}\right), \\
\tau &=& \bhat\bdot\nabla\btimes\bhat \;=\;  \frac{1}{\beta^{2}} \left( g \;+\; \frac{1}{hq}\right), \label{eq:tau_tok} \\
\bhat^{\prime} &=& -\,\frac{\beta^{\prime}}{\beta}\,\bhat \;-\; \left( \varphi^{\prime}\cos\vartheta + \frac{r}{q}\,\vartheta^{\prime}\right)\frac{\wh{\sf r}}{\beta}  \;+\; \frac{gr^{\prime}}{\beta}\;\wh{\vartheta}  \nonumber \\
 &&+\;  \frac{\varphi^{\prime}\sin\vartheta}{\beta}\left( \wh{\vartheta} \;-\; \frac{r}{q}\,\wh{\varphi}\right),
\end{eqnarray}
where $g(r) \equiv d(r/q)/dr$ and $\beta^{\prime} = (rg/q\beta)\,r^{\prime}$.

\subsection{Particle dynamics}

 The  dimensionless particle Lagrangian is
 \begin{equation}
 L = \frac{1}{\epsilon} \left( \Psi \,\vartheta ' \:-\: \psi \,\varphi ' \right) \;+\; \frac{1}{2} \left( r^{\prime 2} + r^2 \, \vartheta^{\prime 2} + h^2 \, \varphi^{\prime 2} \right),
 \end{equation}
from which we obtain the following equations of motion
\begin{eqnarray}
r'' &=& \frac{r}{\epsilon}\left(\frac{\vartheta^{\prime}}{h} - \frac{\varphi^{\prime}}{q}\right) + r \,\vartheta^{\prime 2} + h \, \cos\vartheta \,\varphi^{\prime 2}, \\
\vartheta '' &=& \frac{- 1}{r} \left( \frac{r'}{\epsilon \, h} + h \,\sin\vartheta \, \varphi^{\prime 2} + 2 \, r' \, \vartheta ' \right), \\
\varphi '' &=& \frac{r \, r'}{\epsilon \, h^2 \, q} - \frac{2 \, \varphi '}{h} \, \left( r' \, \cos\vartheta - r \, \vartheta ' \, \sin\vartheta \right).
\end{eqnarray}
These dimensionless equations of motion are solved numerically for $\epsilon = 1/100$, with the initial conditions $(r_{0},\vartheta_{0},\varphi_{0}) = (1/2,0,0)$ and $(r_{0}^{\prime},\vartheta_{0}^{\prime},\varphi_{0}^{\prime}) = (0,8,\sqrt{22}/3)$ associated with a dimensionless energy $E = 43/2$.

We note that, because of the azimuthal symmetry of the magnetic field \eqref{eq:B_tokamak}, i.e., the components $(B_{\vartheta}, B_{\varphi})$ are independent of the azimuthal angle $\varphi$, the azimuthal canonical angular momentum  \eqref{eq:P_varphi}, expressed in dimensionless form as
\begin{equation} 
P_{\varphi} \;=\; - \, \frac{1}{\epsilon} \, \psi(r) \: + \: h^{2}(r,\vartheta) \, \varphi',
\label{eq:P_thetatokamak}
\end{equation}
is a constant of the motion. 

\begin{figure}
\epsfysize=2in
\epsfbox{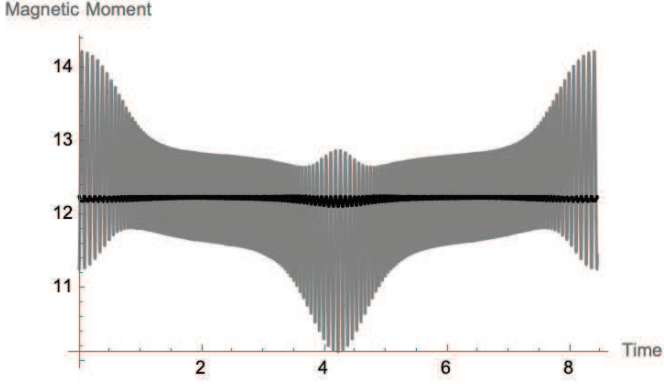}
\caption{Plots of the normalized lowest-order magnetic moment $\mu_{0}$ (gray) and the normalized magnetic moment $\mu = \mu_{0} + \epsilon\,\mu_{1} \simeq 12.23$ (black) for one bounce period, with first-order correction $\mu_{1}$ given by Eq.~\eqref{eq:mu1_tokamak}. Here, $\epsilon = 1/100$ and the normalized kinetic energy is $E_{0} = 21.5$, with the initial conditions $(r_{0},\vartheta_{0},\varphi_{0}) = (1/2,0,0)$ and $(r_{0}^{\prime},\vartheta_{0}^{\prime},\varphi_{0}^{\prime}) = 
(0,8,\sqrt{22}/3)$.}
\label{fig:mu_tokamak}
\end{figure}

Before moving on to guiding center theory, we calculate the normalized magnetic moment $\mu = \mu_0 + \epsilon \,\mu_1  + \cdots$, where
\[ \mu_0 = \left( r^{\prime 2} + r^{2}\omega^{2} \right)/B, \]
with
\begin{equation}
\omega \;\equiv\;  \left( \vartheta^{\prime} \;-\; h\,\varphi^{\prime}/q\right)/\beta,
\label{eq:omega_tok}
\end{equation}
and the dimensionless lowest-order gyroradius is
\begin{equation}
\vb{\rho}_{0} = -\,\frac{r\omega}{B}\,\wh{\sf r} \;+\; \frac{r^{\prime}}{\beta B}\,\left(\wh{\vartheta} - \frac{r}{q}\wh{\varphi}\right).
\end{equation}
Figure \ref{fig:mu_tokamak} shows that the lowest-order normalized magnetic moment is poorly conserved, especially as the particle crosses the equatorial plane $(\vartheta = 0)$. We can greatly improve the adiabatic invariance of the magnetic moment by calculating $\mu_1$ from Eq.~\eqref{eq:mu_1}:
\begin{equation}
\mu_1 = \mu_0 \, G_1 + \frac{p_{\parallel} ^2}{B} \, G_2 + \frac{p_{\parallel}}{B} \,  G_3 - \frac{3 \, \mu_0 \, p_{\parallel}}{2 \,B} \, \tau,
\label{eq:mu1_tokamak}
\end{equation}
where 
\begin{eqnarray} 
G_1 &=& \vb{\rho}_0 \bdot \nabla \ln B = \frac{r^{\prime}}{\beta^{2}}\,\sin\vartheta - \frac{r\omega}{B} \left(\frac{r\,g}{q\beta^{2}} - \frac{\cos\vartheta}{h}\right), \label{eq:G1_tok} \\
G_2 &=& \vb{\rho}_0 \bdot\vb{\kappa} = \frac{r^{\prime}}{\beta^{2}}\,\sin\vartheta + \frac{r\omega}{\beta^{2}B}\left(\frac{\cos\vartheta}{h} + \frac{r}{q^{2}}\right),  \label{eq:G2_tok}
\end{eqnarray}
and
\begin{eqnarray}
G_3 \;=\; \vb{\rho}_{0}\bdot\bhat^{\prime} &=& \frac{r^{\prime}}{B}\left( \frac{g\,r^{\prime}}{\beta^{2}} + \varphi^{\prime}\sin\vartheta \right) \nonumber \\
 &&+\; \frac{r\omega}{\beta B}\left( \frac{r\vartheta^{\prime}}{q} + \varphi^{\prime}\cos\vartheta\right).  \label{eq:G3_tok}
\end{eqnarray}
Figure \ref{fig:mu_tokamak} shows lowest order magnetic moment $\mu_0$ and the improved $\mu = \mu_0 + \epsilon \,\mu_1$.  This first-order correction causes $\mu \simeq 12.23$ to be a good adiabatic invariant, numerically validating the guiding-center representation for the particle orbits in the simple tokamak magnetic field \eqref{eq:B_tokamak}.  This adiabatic invariance of the magnetic moment justifies our use of the guiding-center approximation in describing particle orbits in simple magnetic tokamak geometry. 

\subsection{Guiding-center dynamics}

The guiding-center Lagrangian in simple tokamak geometry is
\begin{equation}
    L_{\rm gc} = \frac{1}{\epsilon} \left( \Psi^* \, \Theta' - \psi^* \, \Phi' \right) - \epsilon\,J \,R' \, {\cal R}_{R}^{*} - \left( \frac{1}{2}\,P_{\|}^2 + J\, B \right),
    \label{eq:Lgc_tokamak}
\end{equation}
where ${\bf X} = (R,\Theta,\Phi)$ denote the guiding-center position in quasi-cylindrical geometry, $J = \mu/2$ denotes the dimensionless guiding-center magnetic moment, and $P_{\|}$ denotes the dimensionless guiding-center momentum parallel to the magnetic field. In addition, the effective poloidal and toroidal magnetic fluxes
\begin{eqnarray}
    \Psi^* &=& \Psi \;+\; \epsilon \, P_{\parallel} \, b_{\Theta} \;-\; \epsilon^2 \,J \,{\cal R}^{*}_{\Theta}, \label{eq:Psi_star_def} \\
    \psi^* &=& \psi \;-\; \epsilon \, P_{\parallel} \, b_{\Phi} \;+\; \epsilon^2 \,J \, {\cal R}^{*}_{\Phi} \label{eq:psi_star_def}
\end{eqnarray}
are expressed in terms of the extended guiding-center model (C) in Eq.~\eqref{eq:R_star_def}: $\vb{\cal R}^{*} = \vb{\cal R} + \frac{1}{2}\,\nabla\btimes\bhat$. Here, we calculate the gyrogauge vector $\vb{\cal R} = \nabla\wh{\sf e}_{1}\bdot\wh{\sf e}_{2}$ by choosing $\wh{\sf e}_{1} = \wh{\sf r}$ and $\wh{\sf e}_{2} = \bhat\btimes\wh{\sf r}$, so that we obtain $\vb{\cal R} = \beta^{-1}\nabla\Theta - b_{z}\nabla\Phi$, where $b_{z} \equiv \bhat\bdot\wh{\sf z} = (R/q\beta)\,\cos\Theta$, and, according to the extended guiding-center model (C) in Eq.~\eqref{eq:R_star_def}, we find
\begin{eqnarray}
    \vb{\cal R}^{*} &=&  \vb{\cal R} \;+\; \frac{1}{2}\left(\nabla b_{\Phi}\btimes\nabla\Phi \;+\frac{}{} \nabla b_{\Theta}\btimes\nabla\Theta\right) \nonumber \\
     &\equiv&{\cal R}_{R}^{*} \, \nabla R \: + \: {\cal{R}}_{\Theta}^* \, \nabla \Theta \: + \: {\cal{R}}_{\Phi}^* \, \nabla \Phi.
\end{eqnarray}
From this Lagrangian, we obtain Euler-Lagrange equations that can be expressed as Eqs.~\eqref{eq:Xgc_dot}-\eqref{eq:Pgc_dot}, where
\begin{equation}
    {\bf B}^{*} = \nabla\Psi^{*}\btimes\nabla\Theta - \nabla\psi^{*}\btimes\nabla\Phi + \nabla\chi\btimes\nabla R, 
    \end{equation}
with $\chi \equiv -\,\epsilon^{2}\,J{\cal R}_{R}^{*}$ and $B_{\|}^{*} = \bhat\bdot{\bf B}^{*}$. The guiding-center equations are thus expressed as
\begin{eqnarray}
R^{\prime} &=& -\;\frac{P_{\|}}{{\cal J}_{\rm gc}}\;\pd{\psi^{*}}{\Theta} \;-\; \frac{\epsilon\,J}{{\cal J}_{\rm gc}}\;b_{\Phi}\pd{B}{\Theta}, \\
\Theta^{\prime} &=&  \frac{P_{\|}}{{\cal J}_{\rm gc}}\;\pd{\psi^{*}}{R} \;+\; \frac{\epsilon\,J}{{\cal J}_{\rm gc}}\;b_{\Phi}\pd{B}{R}, \\
\Phi^{\prime} &=& \frac{P_{\|}}{{\cal J}_{\rm gc}}\left(\pd{\Psi^{*}}{R} + \epsilon^{2}J\;\pd{{\cal R}_{R}^{*}}{\Theta}\right) \;-\; \frac{\epsilon\,J}{{\cal J}_{\rm gc}}\;b_{\Theta}\pd{B}{R}, \\
P_{\|}^{\prime} &=& -\;\frac{J}{{\cal J}_{\rm gc}}\left( \pd{\psi^{*}}{R}\;\pd{\ov B}{\Theta} \;-\; \pd{\psi^{*}}{\Theta}\;\pd{B}{R}\right),
\end{eqnarray}
where ${\cal J}_{\rm gc} \equiv {\cal J}\,B_{\|}^{*}$ combines the quasi-cylindrical Jacobian ${\cal J}$ and the guiding-center Jacobian $B_{\|}^{*}$. We note that these equations exactly conserve the guiding-center energy ${\cal E} = P_{\|}^{2}/2 + J B$ and the guiding-center canonical azimuthal angular momentum 
\begin{equation}
P_{{\rm gc}\Phi} \;=\; -\;\frac{1}{\epsilon}\;\psi^{*} \;=\; -\,\frac{1}{\epsilon}\;\psi \;+\; P_{\|}\,b_{\Phi} \;-\; \epsilon\,J\,{\cal R}_{\Phi}^{*}.
\label{eq:Pgcphi_tok}
\end{equation}
The dimensionless guiding-center equations are solved numerically for $\epsilon = 1/100$ with the initial conditions $(R_{0},\Theta_{0},\Phi_{0}) = (0.5380,0,0)$, which takes into account the radial shift \eqref{eq:initial_X} from the initial particle position $(r_{0},\varphi_{0},z_{0}) = (0.5,0,0)$, and $P_{\|0} = \sqrt{E - J\,B(R_{0},\Theta_{0})}$.

\begin{figure}
\epsfysize=1.8in
\epsfbox{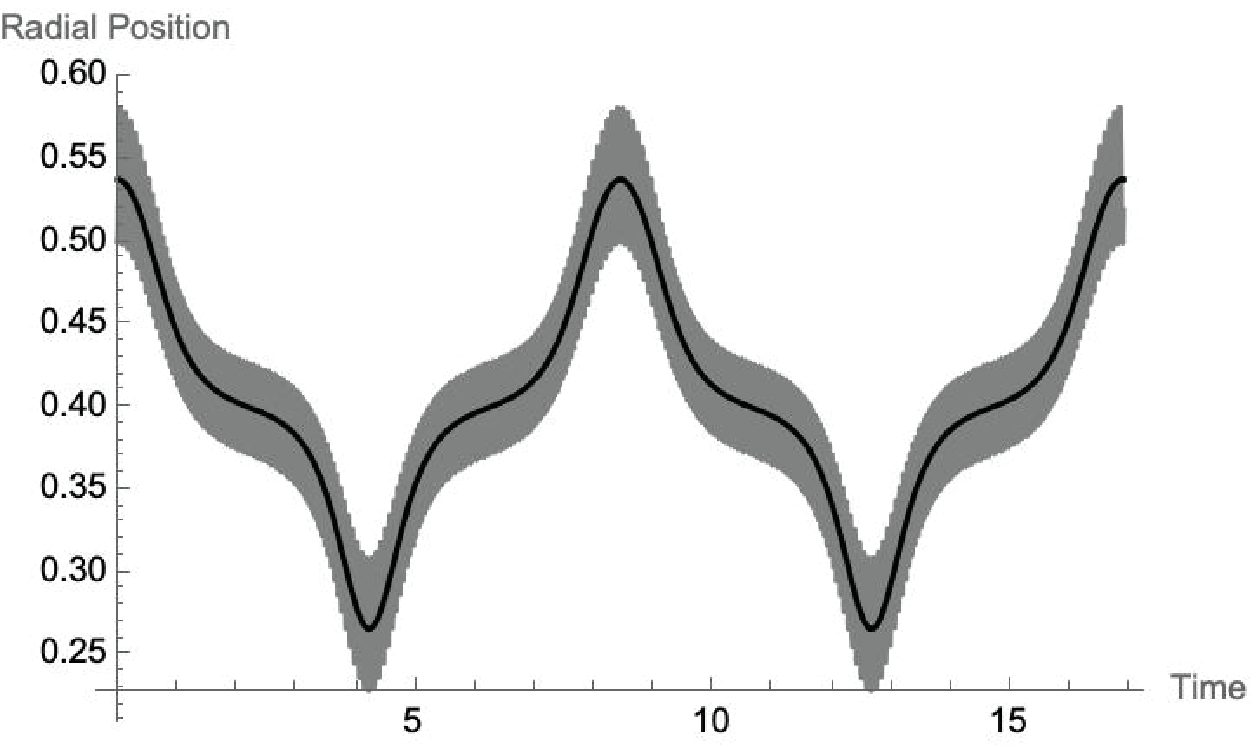}
\caption{Plots of the particle radial position (gray) and the guiding-center radial position (black) during the first two bounce periods.}
\label{fig:r_Tokamak}
\end{figure}

\begin{figure}
\epsfysize=2in
\epsfbox{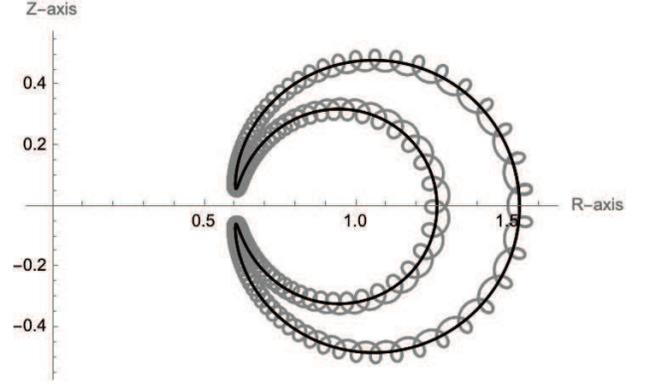}
\caption{Plots of the particle  position (gray) and the guiding-center position (black) projected into the poloidal plane during the first two bounce periods.}
\label{fig:Banana_Tokamak}
\end{figure}

\begin{figure}
\epsfysize=2.6in
\epsfbox{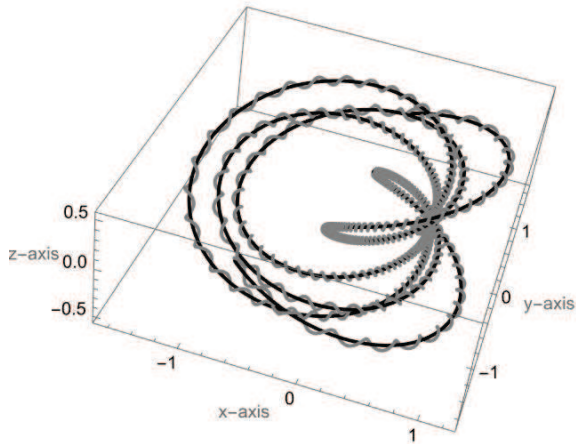}
\caption{Plots of the three-dimensional particle orbit (gray) and the guiding-center orbit (black) during the first two bounce periods.}
\label{fig:3d_Tokamak}
\end{figure}

Figures \ref{fig:r_Tokamak}-\ref{fig:3d_Tokamak} show plots of the particle position (gray) and the guiding-center position (black), obtained from the extended guiding-center model C in Eq.~\eqref{eq:R_star_def}, during the first bounce periods.
Figure \ref{fig:r_Tokamak} shows plots of the particle radial position (gray) and the guiding-center radial position (black) during the first two bounce periods, Fig.~\ref{fig:Banana_Tokamak} shows the classic ``closed'' guiding-center banana orbit (black) projected onto the poloidal plane (at constant toroidal angle), and Fig.~\ref{fig:3d_Tokamak} shows that the three-dimensional guiding-center orbit (black) follows the three-dimensional particle orbit (gray) very well over two bounce periods.

\subsection{Higher-order guiding-center orbits}

\begin{figure}
\epsfysize=2.6in
\epsfbox{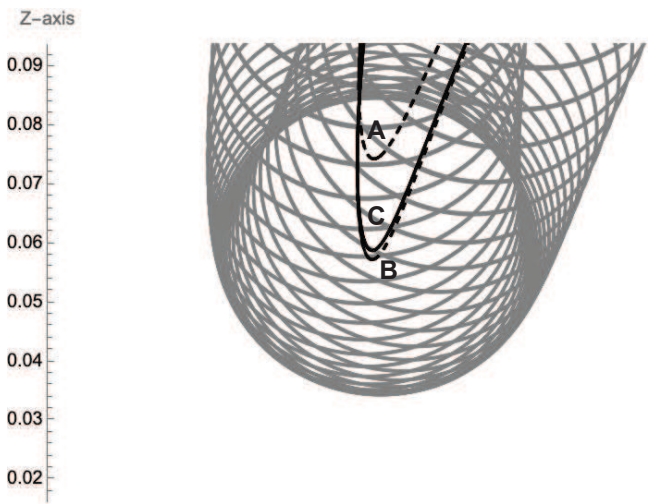}
\caption{Plots of the particle  position (gray) and the guiding-center positions projected into the poloidal plane near the upper turning point, according to three different guiding-center models in Eq.~\eqref{eq:R_star_def}: (black, solid) Extended guiding-center model (C); (black, dashed) Standard guiding-center model (B); and (black, dotted) Truncated guiding-center model (A).}
\label{fig:bounce_Tokamak_models}
\end{figure}

We now make a few remarks on the three guiding-center models presented in Eq.~\eqref{eq:R_star_def}. In order to compare their effectiveness at approximating the particle orbit, we need to ensure that the initial conditions for these guiding-center orbits are consistent with the initial conditions for the particle orbit. This consistency is achieved by connecting the initial conditions through the  guiding-center transformation \eqref{eq:initial_X}. 

In what follows, the guiding-center orbit A is generated from the initial condition obtained from the lowest-order relation ${\bf X} = {\bf x} - \epsilon\,\vb{\rho}_{0}$ (i.e., the initial condition only takes into account the lowest-order guiding-center transformation), which yields the initial radial position $R_{0}|_{A} = 0.5396$ from the particle initial conditions. The initial conditions for the guiding-center orbits B and C, on the other hand, are distinguished by the guiding-center polarization correction $G_{2}^{\bf x}|_{\rm pol} = (J/2m\Omega)\,\vb{\kappa}$ for the extended guiding-center model C \cite{Tronko_Brizard:2015}, while $G_{2}^{\bf x}|_{\rm pol} = 0$ for the standard guiding-center model B \cite{Littlejohn:1983,Brizard:1989}. Hence, we use the initial radial positions $R_{0}|_{C} = 0.5380$ and $R_{0}|_{B} = 0.5384$, which are calculated when the particle initial conditions are inserted in the guiding-center transformation \eqref{eq:initial_X}.

Figure \ref{fig:bounce_Tokamak_models} shows the particle position (gray) and the guiding-center positions (labeled A, B, and C) projected into the poloidal plane near the upper turning point. Because the guiding-center orbits have different guiding-center toroidal canonical angular momenta \eqref{eq:psi_star_def}, Fig.~\ref{fig:bounce_Tokamak_models} shows that each guiding-center turning point occurs on a different magnetic surface. In addition, while the standard guiding-center orbit B slightly overshoots the particle center of gyration at the turning point, the truncated guiding-center orbit A largely undershoots the particle center of gyration. We note that the three guiding-center orbits are nearly indistinguishable away from the turning-point regions (see Fig.~\ref{fig:Banana_Tokamak}).

\begin{figure}
\epsfysize=2.2in
\epsfbox{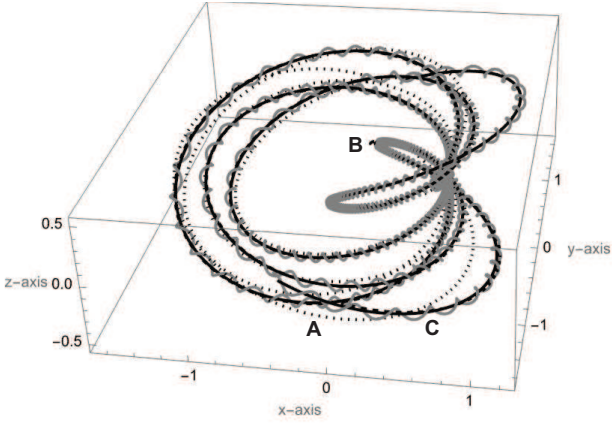}
\caption{Plots of the three-dimensional particle orbit (gray) and the guiding-center orbits (labeled A, B, and C) during two bounce periods, according to three different guiding-center models in Eq.~\eqref{eq:R_star_def}: (black, solid) Extended guiding-center model (C); (black, dashed) Standard guiding-center model (B); and (black, dotted) Truncated guiding-center model (A). Here, the truncated guiding-center orbit A has clearly separated from the particle orbit, while the higher-order guiding-center orbits B and C are nearly indistinguishable, except when the standard guiding-center orbit B overshoots the turning point.}
\label{fig:3d_Tokamak_models}
\end{figure}

Figure \ref{fig:3d_Tokamak_models} shows the three-dimensional particle orbit (gray) and three guiding-center orbits (labeled A, B, and C) during the first two bounce periods. Here, we see that the lowest-order truncated guiding-center orbit A clearly separates from the particle orbit (i.e., the guiding-center orbit is located well outside of the particle's gyration radius), while the two higher-order guiding-center orbits B and C are still nearly indistinguishable over the first two bounce periods, except when the standard guiding-center orbit B overshoots the turning point. We note that the standard guiding-center orbit B noticeably separates from the particle orbit during the next two bounce periods, while the extended guiding-center orbit C still remains at the center of the particle orbit.

\subsection{Validity of the guiding-center representation in simple tokamak magnetic geometry}

We now show that the guiding-center approximation is faithful to the particle motion in simple tokamak magnetic geometry by showing that the guiding-center pull-back ${\sf T}_{\rm gc}P_{{\rm gc}\Phi} = P_{\varphi}$ of the guiding-center canonical azimuthal angular momentum is equal to the particle canonical azimuthal angular momentum. Up to second order in $\epsilon$, the guiding-center pull-back ${\sf T}_{\rm gc}P_{{\rm gc}\Phi}$ is expressed as
\begin{eqnarray}
{\sf T}_{\rm gc}P_{{\rm gc}\Phi} &=& -\frac{1}{\epsilon}\,\psi \;+\; \vb{\rho}_{0}\bdot\nabla\psi \;+\; p_{\|}\;b_{\varphi} 
 \label{eq:Tgc_Pphi} \\
 &&+\; \epsilon\;\vb{\rho}_{1}\bdot\nabla\psi \;-\; \frac{\epsilon}{2}\,\vb{\rho}_{0}\vb{\rho}_{0}:\nabla\nabla\psi \nonumber \\
 &&+\; \epsilon \left(G_{1}^{p_{\|}}\,b_{\varphi} - p_{\|}\,\vb{\rho}_{0}\bdot\nabla b_{\varphi}\right) \;-\; \epsilon\,J\,{\cal R}_{\varphi}^{*},
\nonumber
\end{eqnarray}
where the first-order gyroradius correction $\vb{\rho}_{1}$ is given by Eq.~\eqref{eq:rho_1} and the first-order correction to the guiding-center parallel momentum is given by Eq.~\eqref{eq:ppar_1}:
\begin{equation}
G_{1}^{p_{\|}} \;=\; -\;\frac{p_{\|}}{2}\,G_{2} \;+\; \frac{3}{2}\,J\;\tau \;-\; \frac{1}{2}\,G_{3},
 \label{eq:ppar1}
 \end{equation}
 where $(\tau,G_{2},G_{3})$ are defined in Eqs.~\eqref{eq:tau_tok} and \eqref{eq:G2_tok}-\eqref{eq:G3_tok}.

 First, using $\vb{\rho}_{0} = \bhat\btimes{\bf x}^{\prime}/B$ and the simple-tokamak identity ${\bf B}\btimes\partial{\bf x}/\partial\varphi = \nabla\psi$, we find
 \[  \vb{\rho}_{0}\bdot\nabla\psi \;=\; \frac{\bhat}{B}\btimes{\bf x}^{\prime}\bdot{\bf B}\btimes\pd{\bf x}{\varphi} \;=\; h^{2}\varphi^{\prime} \;-\; p_{\|}\,b_{\varphi}, \]
 so that, at first order (i.e., zeroth order in magnetic-field nonuniformity), we find
 \begin{equation}
 \vb{\rho}_{0}\bdot\nabla\psi \;+\; p_{\|}\;b_{\varphi} \;=\; h^{2}\,\varphi^{\prime}.
 \end{equation}
Hence, we now need to show that, at second order (i.e., first order in magnetic-field nonuniformity), we find the identity
 \begin{eqnarray}
J{\cal R}_{\varphi}^{*} &\equiv& \vb{\rho}_{1}\bdot\nabla\psi \;-\; \frac{1}{2}\vb{\rho}_{0}\vb{\rho}_{0}:\nabla\nabla\psi \nonumber \\
 &&+\; G_{1}^{p_{\|}}b_{\varphi} \;-\; p_{\|}\vb{\rho}_{0}\bdot\nabla b_{\varphi},
\label{eq:ang_mom_id}
 \end{eqnarray}
 so that Eq.~\eqref{eq:Tgc_Pphi} becomes
 \begin{equation}
{\sf T}_{\rm gc}P_{{\rm gc}\Phi} \;=\;  -\frac{1}{\epsilon}\,\psi \;+\; h^{2}\,\varphi^{\prime} \;\equiv\; P_{\varphi},
\end{equation}
which guarantees the validity of the guiding-center representation in simple tokamak magnetic geometry. A complete proof of the identity \eqref{eq:ang_mom_id} is given in App.~\ref{sec:proof}.

\section{Summary}

In previous work \cite{Brizard:2017}, we showed that the guiding-center approximation was valid in a straight magnetic field with constant perpendicular magnetic gradient, even in the presence of strong gradients. In addition, based on the existence of an exact analytical solution for the particle orbits, this work also confirmed that the guiding-center polarization corresponded exactly with an orbit-averaged particle displacement.

In the present work, we extended our investigation of the validity of the guiding-center approximation in describing charged single-particle motion in a nonuniform magnetic field. Here, we considered regular particle orbits in azimuthally symmetric magnetic mirror geometry and simple tokamak magnetic geometry, in which the azimuthal angular canonical momentum is conserved and the guiding-center magnetic moment is an adiabatic invariant. We successfully validated the guiding-center approximation in describing particle motion in an azimuthally symmetric magnetic field provided higher-order guiding-center corrections are taken into account, which had already been noted for the case of an axisymmetric tokamak magnetic field \cite{Belova:2003}. In particular, the guiding-center polarization correction in the guiding-center azimuthal angular canonical momentum, not taken into account in the standard guiding-center approximation \cite{Littlejohn:1983,Brizard:1989}, proved crucial in establishing a faithful guiding-center representation for regular particle orbits in axisymmetric magnetic geometry.

Lastly, we note that the truncated guiding-center model \cite{White:2014} is used extensively in guiding-center particle simulations, despite the fact that it is not as faithful to particle orbits as higher-order guiding-center models. In most applications, however, the truncated guiding-center model is either used to analyze particle orbits in the presence of perturbed electric and/or magnetic fields, or as the unperturbed component for the gyrocenter orbit used in nonlinear gyrokinetic theory \cite{Brizard:1989}. Future work may look into the issue of faithfulness for these applications. In addition, the faithfulness of the guiding-center representation for particle orbits in non-axisymmetric magnetic geometries may be explored.

\appendix

\section{\label{sec:gc}Guiding-center Transformation}

The standard expression for the first-order correction to the guiding-center magnetic moment is \cite{Littlejohn:1983,Tronko_Brizard:2015}
\begin{equation}
G_{1}^{\mu} = \vb{\rho}_{0}\bdot\left(\mu_{0}\,\nabla\ln B + \frac{p_{\|}^{2}\,\vb{\kappa}}{m\,B}\right) - \mu_{0}\;\varrho_{\|}\,(\tau + \alpha_{1}),
 \label{eq:mu1_RGL}
 \end{equation}
 where $\varrho_{\|} \equiv p_{\|}/(m\Omega)$ and $\alpha_{1} \equiv -\,\frac{1}{2}\,(\wh{\bot}\wh{\rho} + \wh{\rho}\wh{\bot}):\nabla\bhat$ is constructed from the gyrorangle-dependent unit vectors $\wh{\bot} \equiv \partial\wh{\rho}/\partial\zeta = \wh{\rho}\btimes\bhat$. Using the identity
 \begin{equation}
 \alpha_{1} \;=\; \frac{1}{2}\,\tau \;-\; \wh{\bot}\bdot\nabla\bhat\bdot\wh{\rho},
 \end{equation}
 we obtain
 \begin{equation}
 \tau + \alpha_{1} \;=\; \frac{3}{2}\,\tau \;-\; \wh{\bot}\bdot\nabla\bhat\bdot\wh{\rho}.
 \label{eq:tau_alpha1}
 \end{equation}
 Next, we write
 \begin{eqnarray} 
 \mu_{0}\;\varrho_{\|}\,\left(\wh{\bot}\bdot\nabla\bhat\bdot\wh{\rho}\right) &=& \frac{p_{\|}}{2B}\;\left({\bf v}_{\bot}\bdot\nabla\bhat\bdot\vb{\rho}_{0}\right) \nonumber \\
  &=& \frac{p_{\|}}{2B}\;\left(\frac{d\bhat}{dt} \;-\; v_{\|}\,\bhat\bdot\nabla\bhat\right)\bdot\vb{\rho}_{0} \nonumber \\
   &=& \left(\frac{p_{\|}}{2B}\;\frac{d\bhat}{dt} \;-\; \frac{p_{\|}^{2}\vb{\kappa}}{2\,mB}\right)\bdot\vb{\rho}_{0},
  \end{eqnarray}
  which yields Eq.~\eqref{eq:mu_1}:
\begin{eqnarray}
G_{1}^{\mu} &=& \left( \mu_{0}\nabla\ln B + \frac{p_{\|}^{2}\,\vb{\kappa}}{2\,mB}\right)
\bdot\vb{\rho}_{0} - \frac{3}{2}\,\mu_{0} \left( \frac{p_{\|}\,\tau}{m\Omega}\right) \nonumber \\
 &&+\; \frac{p_{\|}}{2B}\;\frac{d\bhat}{dt}\bdot\vb{\rho}_{0}.
\end{eqnarray}

Using the same identity \eqref{eq:tau_alpha1}, the standard expression for the first-order correction to the guiding-center parallel momentum \cite{Littlejohn:1983,Tronko_Brizard:2015} is replaced with the new expression
\begin{eqnarray}
G_{1}^{p_{\|}} &=& -\;p_{\|}\,\vb{\rho}_{0}\bdot\vb{\kappa} \;+\; \frac{\mu_{0}B}{\Omega}\;(\tau + \alpha_{1}) \nonumber \\
 &=& -\;\frac{p_{\|}}{2}\,\vb{\rho}_{0}\bdot\vb{\kappa} \;+\; \frac{3}{2}\,\frac{\mu_{0}B}{\Omega}\;\tau \;-\; \frac{m}{2}\,\frac{d\bhat}{dt}\bdot\vb{\rho}_{0}.
 \label{eq:ppar_1}
 \end{eqnarray}
 
 \section{\label{sec:gc_pol}Guiding-center Polarization}
 
 The guiding-center polarization was calculated directly from the guiding-center transformation in our previous works \cite{Brizard:2013,Tronko_Brizard:2015}. It is formally defined by the multipole expansion
 \begin{equation}
 \vb{\pi}_{\rm gc} \;\equiv\; e\,\langle\vb{\rho}_{\rm gc}\rangle \;-\; \nabla\bdot\left(\frac{e}{2}\;\langle\vb{\rho}_{\rm gc}\vb{\rho}_{\rm gc}\rangle\right) \;+\; \cdots,
 \label{eq:pi_gc}
 \end{equation}
 where the dipole and quadrupole moments are shown here, while $\vb{\rho}_{\rm gc} \equiv {\sf T}_{\rm gc}^{-1}{\bf x} - {\bf X}$ is the guiding-center gyroradius. We note that the guiding-center gyroradius $\vb{\rho}_{\rm gc}$ is related to the particle gyroradius $\vb{\rho} \equiv {\bf x} - {\sf T}_{\rm gc}{\bf X}$ by the identity $\vb{\rho}_{\rm gc} \equiv {\sf T}_{\rm gc}^{-1}\vb{\rho}$. Using the guiding-center transformation presented by Tronko and Brizard \cite{Tronko_Brizard:2015}, we find the dipole contribution
 \begin{eqnarray}
\langle\vb{\rho}_{\rm gc}\rangle &=& \epsilon^{2} \left( \langle\vb{\rho}_{1}\rangle - \frac{p_{\|}^{2}\vb{\kappa}}{m^{2}\Omega^{2}} \right) \nonumber \\
 &&-\; \frac{\epsilon^{2}\mu B}{m\Omega^{2}} \left[ 2\,\nabla_{\bot}\ln B \;+\frac{}{} \left(\nabla\bdot\bhat\right)\bhat \right],
 \end{eqnarray}
 and the quadrupole contribution
 \begin{eqnarray} 
-\, \nabla\bdot\left(\frac{1}{2}\;\langle\vb{\rho}_{\rm gc}\vb{\rho}_{\rm gc}\rangle\right) &=& -\,\nabla\bdot\left( \frac{\epsilon^{2}\mu B}{2m\Omega^{2}}\;(\mathbb{I} - \bhat\bhat)\right) \\
  &=& \frac{\epsilon^{2}\mu B}{2m\Omega^{2}}\;\nabla_{\bot}\ln B \nonumber \\
   &&+\; \frac{\epsilon^{2}\mu B}{2m\Omega^{2}} \left[ \vb{\kappa} \;+\frac{}{} \left(\nabla\bdot\bhat\right)\bhat \right],  \nonumber
 \end{eqnarray}
 which both appear at $\epsilon^{2}$ at their lowest orders  (i.e., first order in magnetic-field nonuniformity).
 
 Hence, the guiding-center polarization \eqref{eq:pi_gc} is expressed as
 \begin{eqnarray}
 \vb{\pi}_{\rm gc} &=& \epsilon^{2}e \left( \langle\vb{\rho}_{1}\rangle_{\rm pol} \;+\; \frac{\mu B}{2\,m\Omega^{2}}\;\vb{\kappa} \right) \nonumber \\
  &&+\; \frac{\epsilon^{2}e\bhat}{\Omega}\btimes\left[\frac{\bhat}{m\Omega}\btimes\left(\mu\,\nabla B + \frac{p_{\|}^{2}}{m}\,\vb{\kappa}\right)\right],
  \end{eqnarray} 
 where $\langle\vb{\rho}_{1}\rangle_{\rm pol}$ is the polarization correction not included in the standard guiding-center transformation \cite{Brizard:1989,Cary_Brizard:2009}. We, therefore, recover the standard guiding-center polarization $\vb{\pi}_{\rm gc} \equiv \epsilon^{2}(e\bhat/\Omega)\btimes d{\bf X}/dt$ \cite{Kaufman:1986} only if we choose 
 \begin{equation}
 \langle\vb{\rho}_{1}\rangle_{\rm pol} \;=\; -\,\frac{\mu B}{2\,m\Omega^{2}}\;\vb{\kappa},
 \label{eq:rho1_pol}
 \end{equation}
 which appears as the first term on the right side of Eq.~\eqref{eq:rho_1}. With this choice, the magnetic vector potential ${\bf A}^{*}$ defined in Eq.~\eqref{eq:Pgc_varphi} becomes
 \begin{eqnarray}
 \frac{e{\bf A}^{*}}{\epsilon c} &=&  \frac{e{\bf A}}{\epsilon c} + P_{\|}\,\bhat \;-\; \epsilon J \left(\vb{\cal R} + \frac{1}{2}\tau\bhat\right) \;+\; \frac{m\Omega}{\epsilon}\, \langle\vb{\rho}_{1}\rangle_{\rm pol} \nonumber \\
  &=& \frac{e{\bf A}}{\epsilon c} + P_{\|}\,\bhat \;-\; \epsilon J \left(\vb{\cal R} + \frac{1}{2}\,\nabla\btimes\bhat\right),
  \end{eqnarray}
  where the standard correction $ \frac{1}{2}\tau\bhat$ \cite{Brizard:1989,Cary_Brizard:2009} is replaced with the correction $\frac{1}{2}\,\nabla\btimes\bhat$ \cite{Brizard:2013,Tronko_Brizard:2015}.
  
  Lastly, it is important to keep in mind that the guiding-center polarization discussed here occurs in the absence of an external electric field and is simply due to magnetic-field non-uniformity. Since the guiding-center polarization effect is inversely proportional the the gyrofrequency, it is an important effect for ions. The electric field generated by the charge separation associated with the guiding-center polarization, therefore, requires a self-consistent treatment that must include an electric field as an integral part of the guiding-center formulation \cite{Brizard:1995}. This self-consistent analysis, however, is outside the scope of this paper.
 
 \section{\label{sec:proof}Proof of Identity \eqref{eq:ang_mom_id}}
 
In this Appendix, we proceed with a proof of the identity \eqref{eq:ang_mom_id}. Using $(\vb{\kappa}/B)\bdot\nabla\psi = \tau\,b_{\varphi} - 2\,\left(b_{z} + {\cal R}_{\varphi}^{*}\right)$, we write
 \begin{eqnarray}
 \vb{\rho}_{1}\bdot\nabla\psi &=& -\,\frac{1}{2}\,J \left[ \tau\,b_{\varphi} \;-\frac{}{} 2\,\left(b_{z} + {\cal R}_{\varphi}^{*}\right)\right] \nonumber \\
   &&+\; \left(\frac{1}{2}\,G_{1} - \frac{p_{\|}}{B}\,\tau\right) \vb{\rho}_{0}\bdot\nabla\psi ,
 \end{eqnarray}
Next, since $\nabla\psi = (r/q)\,\wh{\sf r}$, we find
\begin{eqnarray} 
\nabla\nabla\psi &=& g\;\wh{\sf r}\,\wh{\sf r} \;+\; \frac{1}{q}\,\wh{\vartheta}\wh{\vartheta} \;+\; B\,b_{z}\;\wh{\varphi}\wh{\varphi} \nonumber \\
 &=& B\,b_{z}\;{\sf I} \;+\; \left( g \;-\; \frac{r\cos\vartheta}{qh}\right)\;\wh{\sf r}\,\wh{\sf r} \;+\; \frac{\wh{\vartheta}\wh{\vartheta}}{qh},
\end{eqnarray}
so that
\begin{equation}
\frac{1}{2}\vb{\rho}_{0}\vb{\rho}_{0}:\nabla\nabla\psi = J b_{z} + \frac{1}{2}\left( g - \frac{r\cos\vartheta}{qh}\right)\rho_{0r}^{2} + \frac{\rho_{0\vartheta}^{2}}{2\,qh},
 \end{equation}
 where $\rho_{0r} = -\,r\omega/B$, $\rho_{0\vartheta} = r^{\prime}/\beta B$, and we used the lowest-order expression
 \[ J \;=\; \frac{B}{2}\left(\rho_{0r}^{2} \;+\frac{}{} \beta^{2}\rho_{0\vartheta}^{2}\right) \;=\; \frac{1}{2B} \left( r^{\prime 2} \;+\; r^{2}\omega^{2}\right), \]
where $\rho_{0\varphi} = -(r/q)\,\rho_{0\vartheta}$. With these expressions, and using Eq.~\eqref{eq:ppar1}, the identity \eqref{eq:ang_mom_id} becomes
 \begin{eqnarray*}
 J\,{\cal R}_{\varphi}^{*} &=& -\,\frac{J}{2} \left[ \tau\,b_{\varphi} - 2\frac{}{}\left(b_{z} + {\cal R}_{\varphi}^{*}\right)\right] \;-\; J\,b_{z} \;+\; \frac{3}{2}\,J\; \tau\,b_{\varphi} \\
  &&+\; \frac{p_{\|}}{B}\,\left( G_{1} \;-\; \frac{1}{2}\,G_{2} \;-\; \tau\;\vb{\rho}_{0}\bdot\nabla\psi \right) \;-\; \frac{G_{3}}{2B} \\
  &&-\; \frac{\rho_{0\vartheta}^{2}}{2\,hq} - \frac{1}{2}\left(g - \frac{r\cos\vartheta}{hq}\right)\rho_{0r}^{2} + \frac{1}{2}\,G_{1}\;\vb{\rho}_{0}\bdot\nabla\psi,
\end{eqnarray*}
which, after cancellations, yields an expression for $G_{3}$:
\begin{eqnarray}
G_{3} &=& 2\,J\,\tau \;+\; p_{\|} \left( 2\,G_{1} \;-\; G_{2} \;-\frac{}{} 2\,\tau\;\vb{\rho}_{0}\bdot\nabla\psi \right) \label{eq:G3_1} \\
 &&+\; G_{1}\,B\vb{\rho}_{0}\bdot\nabla\psi \;-\; \frac{B\,\rho_{0\vartheta}^{2}}{hq} \;-\; \left(\beta^{2}\,\tau - \frac{1}{q}\right) B\rho_{0r}^{2}, \nonumber 
\end{eqnarray}
where we used
\[ g - \frac{r\cos\vartheta}{hq} \;=\; g + \frac{1}{q}\left(\frac{1}{h} - 1\right) \;=\; \beta^{2}\,\tau - \frac{1}{q}, \]
which follows from the definition \eqref{eq:tau_tok} for $\tau$.
 
We now compare Eq.~\eqref{eq:G3_1} with Eq.~\eqref{eq:G3_tok}, which requires expressing $(\vartheta^{\prime},\varphi^{\prime})$ in terms of $(p_{\|},\omega)$, where $\omega$ is defined in Eq.~\eqref{eq:omega_tok} and $p_{\|} = \varphi^{\prime}/B + r^{2}\vartheta^{\prime}/(q\beta)$ is the lowest-order dimensionless particle parallel momentum. Hence, after substituting
 \begin{eqnarray}
 \vartheta^{\prime} &=& (\omega + p_{\|}/q)/\beta, \nonumber \\
  && \\
 \varphi^{\prime} &=& (p_{\|} - r^{2}\omega/q)/(h\beta), \nonumber
 \end{eqnarray}
 into Eq.~\eqref{eq:G3_tok}, we obtain a second expression for $G_{3}$:
 \begin{eqnarray}
 G_{3} &=& 2\,J\,\tau \;+\; p_{\|}\,G_{2} \;+\; G_{1}\,B\vb{\rho}_{0}\bdot\nabla\psi \label{eq:G3_2} \\
  &&-\; \frac{2J}{hq\,\beta^{2}} \;-\; \frac{r^{2}\omega^{2}}{B} \left[ g \;-\; \frac{(h\beta^{2} + 1 - \beta^{2})}{hq\,\beta^{2}} \right], \nonumber 
 \end{eqnarray}
 where we used $1 + \beta^{2}r\cos\vartheta = h\beta^{2} + 1 - \beta^{2}$. By comparing Eqs.~\eqref{eq:G3_1} and \eqref{eq:G3_2}, we obtain the following expression
 \begin{equation}
p_{\|} \left( G_{1} \;-\; G_{2} \;-\frac{}{} \tau\;\vb{\rho}_{0}\bdot\nabla\psi \right) \;=\; 0,
\label{eq:p_proof}
\end{equation}
after carrying out several cancellations on the right side of Eq.~\eqref{eq:p_proof}. Lastly, using Eqs.~\eqref{eq:G1_tok}-\eqref{eq:G2_tok}, we obtain
\begin{eqnarray}
G_{1} - G_{2} &=& -\,\frac{r\omega}{B} \left( \frac{rg}{q\beta^{2}} - \frac{\cos\vartheta}{h}\right) - \frac{r\omega}{B\beta^{2}} \left( \frac{\cos\vartheta}{h} - \frac{r}{q^{2}}\right) \nonumber \\
 &=& -\,\tau\;\frac{r^{2}\omega}{qB} \;\equiv\; \tau\,\vb{\rho}_{0}\bdot\nabla\psi,
\end{eqnarray}
which confirms Eq.~\eqref{eq:p_proof} and completes the proof of the identity \eqref{eq:ang_mom_id}.

\acknowledgments

The present work was supported by the National Science Foundation grant PHY-2206302.

\vspace*{0.1in}

\begin{center}
{\bf Data Availability Statement}
\end{center}

The Mathematica code used to generate the plots in the present manuscript is available upon request.

\bibliography{GC}

\end{document}